# On the Role of Infrastructure sharing for Mobile Network Operators in Emerging Markets

Djamal-Eddine Meddour[1], Tinku Rasheed[2] and Yvon Gourhant[1]

[1]*France Telecom-Orange R&D, Lannion, France*
[2]*CREATE-NET Research Center, Trento, Italy*

**Abstract**

The traditional model of single ownership of all the physical network elements and network layers by mobile network operators is beginning to be challenged. This has been attributed to the rapid and complex technology migration compounded with rigorous regulatory requirements and ever increasing capital expenditures. These trends, combined together with the increasing competition, rapid commoditization of telecommunication equipments and rising separation of network and service provisioning are pushing the operators to adopt multiple strategies, with network infrastructure sharing in the core and radio access networks emerging as a more radical mechanism to substantially and sustainably improve network costs. Through infrastructure sharing, developing countries and other emerging economies can harness the technological, market and regulatory developments that have fostered affordable access to mobile and broadband services. Similarly, the network operators entering or consolidating in the emerging markets can aim for substantial savings on capital and operating expenses. The present paper aims to investigate the current technological solutions and regulatory and the technical-economical dimensions in connection with the sharing of mobile telecommunication networks in emerging countries. We analyze the estimated savings on capital and operating expenses, while assessing the technical constraints, applicability and benefits of the network sharing solutions in an emerging market context.

## 1. Introduction

Mobile telecommunication services have shown impressive uptake in the past decade. In particular in developing countries, mobile telephony has played a vital role in making cellular services available to a part of the population that did not have access to such services previously. However, considerable advances are required to increase the penetration of mobile services and to improve competition in the cellular market, in particular in rural areas in developing countries. The roll-out of mobile networks requires high sunk investments and the need to recover those by charging the user heavily for accessing mobile services [1]. This often makes mobile services less affordable and may discourage operators to innovate and migrate to new technologies in emerging markets. It may also cause licensed mobile network operators (MNO) to obstruct the entry of new operators in the market and additionally, it may be too costly for new entrant operators to rollout mobile networks in rural and less populated areas, resulting in exclusion of a part of the population or certain regions from access to mobile telecommunication services [3].

Traditional mobile network operation strategy is characterized by a high degree of vertical integration where the MNO acquires and develops the sites needed for rolling out the network, plans the network architecture and topology, operates and maintains the network and customer relationships, creates, markets and provides services to its end users. However, technology migration, such as the introduction of third generation (3G) and 3.5G wireless technologies on top of 2G networks, and the introduction of 4G technologies including LTE, is becoming increasingly rapid and complex [8]. Regulatory requirements also mandate coverage of areas that is not attractive from a business perspective. With growing competitive intensity and rapid price declines, mobile operators are facing increased margin pressure and the need to systematically improve their cost position.

In current market environment, focusing merely on the provisioning of coverage and capacity has a relatively low success factor, and to address this reality, operators are adopting multiple strategies, with network sharing emerging as a more radical mechanism to substantially and sustainably improve network costs. Mobile infrastructure sharing in telecom is an important measure to reduce costs. It is useful in start up phase to build coverage quickly and in the longer term scenario to build more cost effective coverage, especially in rural and less populated or marginalized areas. In the emerging market context, both in urban and rural areas infrastructure sharing should be adopted as an imperative for sustained telecom growth.

Mobile infrastructure sharing may also stimulate the migration to new technologies and the deployment of mobile broadband, which is increasingly seen as a viable means of making broadband services accessible for a larger part of the world population [13]. Mobile sharing may also enhance competition between mobile operators and service providers, at least where certain safeguards are used, without which concerns of anticompetitive behaviour could arise. Ultimately, mobile network sharing can play an important role in increasing access to information and communication technologies (ICTs), generating economic growth, improving quality of life and helping developing and developed countries to meet the objectives established by the World Summit on the Information Society (WSIS) and the Millennium Development Goals established by the United Nations [20].

Different forms of infrastructure sharing are possible, ranging from basic unbundling and national roaming, to advanced forms like collocation and spectrum sharing. In the MENA (Middle East and North Africa) region, National roaming is used extensively in countries like Jordan, Morocco, Oman, Saudi Arabia and the United Arab Emirates. Unbundling is now starting to gather pace, with Egypt and Saudi Arabia as leaders. Other forms of sharing are bound to develop, given the expected returns to incumbents and new entrants alike. While infrastructure sharing is the most cost-efficient design principle for any new roll-out in emerging markets and the best approach for technology migration and consolidation [12], the cost savings potential from infrastructure sharing are earned through sacrificing some of the control that the standalone operator has over its network, thereby impacting the ability of operators to compete and differentiate themselves based on network quality. This is why, considering both the appeal of sharing to the operators, and their strategic interests, the stronger forms of sharing are usually recommended for coverage-driven roll-outs in rural areas that have limited business potential, and where differentiation (which requires autonomy) is less important. In addition, obligations relating to network sharing may influence the willingness of operators to make efficient investments in infrastructure and innovative services.

Network sharing ideas and proposals for different approaches started to appear after the UMTS licenses were granted in Europe in the 2000s [7]. Although both academia and industry have contributed relevant ideas and focus directions, most of the works focus on one or few single aspects of infrastructure sharing. The 3GPP has prepared a technical report on the various models used for sharing [2]. In 2001, TIA Europe drafted a report on the state of shared 3G network infrastructure in Europe [3]. It introduced the places where the infrastructure sharing started in Europe.

In 2009, the NorthStream report [4][9] analyzed the competitive and practical effects of network sharing. The authors in [19] proposed a technical resource sharing framework tailored for the MNO-MVNO-context with an emphasis on service level agreements (SLA). The 3GPP is also working on facilitating the sharing of future networks [5][6]. Contributions from industry typically focussed on the description of off-the shelf technical solutions, but fail to study the operator's processes and to quantify economic implications. Also, given the vendor perspective, most of the attention goes to fixed assets rather than operational considerations which address investment, coverage and time-to-market issues for new roll-outs.

Network operators in developing markets, especially in the MENA region, are growing in number and size significantly. The liberalization trend, coupled with infrastructure sharing prospects offers a potential opportunity for all the stakeholders in each market. Incumbent operators can optimize the use of their existing infrastructure and generate new revenue that can be channelled to international expansions. On the other hand, new entrants leveraging infrastructure sharing will be able to focus on service offerings, with limited rollout burdens. In the liberalized fixed markets, growth and success rely extensively on sharing the incumbent operators' local loop, given the difficulty to rollout competing access networks. As a specific example, market reports indicate that since local loop unbundling was enforced in Morocco earlier this year, the broadband market grew 19% in a period of 6 months [22].

Going forward, and as discussions about the commoditization of telecom networks are materializing, with the separation of network and service provisioning on the rise, infrastructure sharing is expected to reach new heights in emerging markets. Incumbents and new entrant operators are expected to go beyond the limited-scope services like national roaming to explore advanced sharing possibilities like duct sharing or joint network rollout. Competing operators' should identify infrastructure synergies that would pave the way for lower costs and better service offerings.

Recent industry trends show higher awareness and readiness towards network sharing, also among incumbent operators. Where emerging/developing market operators are looking at economic option for coverage and capacity growth, operators in mature markets are seeking cost optimization and technology refresh, by a step further by establishing a joint venture currently aiming at optimizing access transmission through sharing leased lines and microwave links. This paper deals with the different levels of network sharing, benefits and disadvantages of each alternative technology sharing and the various technical and regulatory constraints linked to the deployment and operation of shared networks.

# 2. Technical approaches for Infrastructure sharing

Network sharing is a very complex process. There are a variety of options that may be considered when assessing the viability of infrastructure sharing. Those options range from the sharing of towers and other infrastructure facilities to sharing an entire mobile network. This section identifies a number of technical options, dividing them into three basic categories: (i) passive sharing, (ii) active sharing and (iii) roaming-based sharing.

Passive sharing refers to the sharing of space in passive infrastructure, such as building premises, sites and masts. Passive sharing is typically a moderate form of network sharing, where there are still separate networks that simply share physical space. Active sharing is a more complex type of sharing, where operators share elements of the active layer of a mobile network, such as antennas, radio nodes, node controllers, backhaul and backbone transmission, as well as elements of the core network (such as switches). Roaming-based sharing in the context of network sharing means that one operator relies on another operator's coverage for a certain, defined footprint on a permanent basis. Operators focusing on emerging/developing markets normally look for economic options for coverage and capacity growth and will be more inclined to passive sharing approaches, where as the operators in mature markets are seeking cost optimizations and new technology options, through active sharing opportunities aiming at optimizing access transmission through sharing leased lines and microwave links. The increasing competition in emerging markets are also driving the operators to choose more viable horizontal partnerships including non-strategic asset outsourcing of operations and services.

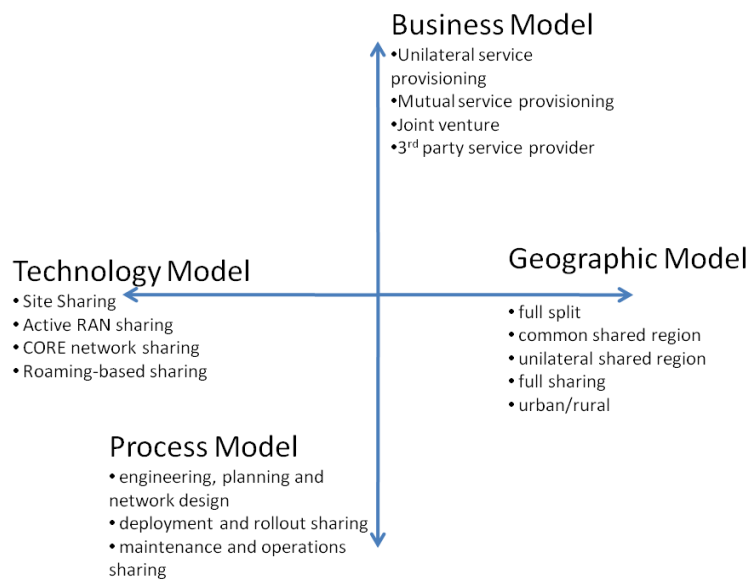

Figure 1: the dynamics of infrastructure sharing

Network sharing can be characterized into four main dimensions, firstly based on the business model, which describes the parties involved and the contractual relationship between the parties; the geographic model, describing each operator's physical footprint; the technology model, describing the technical approach used for sharing (see Figure 1); and the process model determining the services to be shared. Thus, the technological approaches chosen for infrastructure sharing is strongly linked to the business model, the geographic consideration and the process model, and the choice of a certain approach will limit the degree of freedom for reasonable choices of the other dimensions.

The network sharing agreement between the involved operators details the commercial, technical, operational and legal conditions of the partnership. Logistically, the possibilities are:

- New network (Greenfield): Both (or more) operators build a new network together (typically 3G networks), ideally when rolling out a new network generation. At the outset, the new shared network infrastructure and operations can be based on the capacity and coverage requirements of both operators. The operators would fund the build-out on a shared basis or according to their expected capacity needs.

- All-in-one network or Buy-in: a buy-in situation arises when one of the sharing operators have already built out the network and look for other operators to share the network. In this case, the second operator would either pay a capacity usage fee or an up-front fee to acquire a share in the network. One challenge in this situation is determining how to agree on potential adjustments and build-outs that would reflect the needs and requirements of the operator who is buying into the existing network.
- Consolidated network: Operators merge their networks and deconstruct the redundant sites. This type of network sharing usually holds significant cost advantages, but it also presents substantial design challenges.

The sharing of infrastructure can be achieved at different levels. Figure 2 shows the five levels of sharing. For ease of presentation of the infrastructure sharing techniques, we assume that each level involves the sharing of the previous level. Naturally, this condition is not always necessary. More complex combinations of different levels of sharing could be implemented by the operators; for example, it is thus theoretically possible to pool the L1 and L3 levels without the pool level L2. However, regulators can prevent operators from adopting certain configurations for sharing. In some countries, operators cannot share specific levels. In Sweden, operators can share at all levels; the only requirement is that each operator covers at least 30 percent of the population with their own network infrastructure.

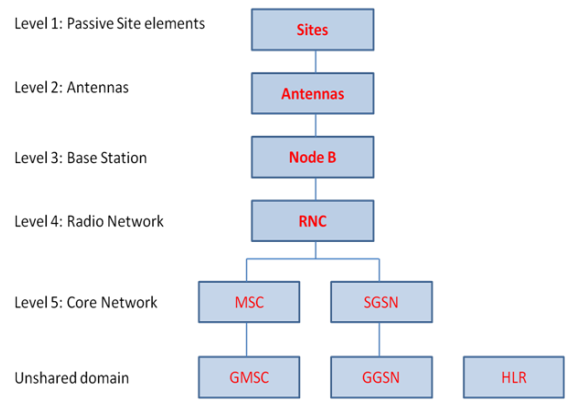

Figure 2: the different levels of infrastructure sharing in mobile networks

It is not necessary to use a single combination of levels across the network. Depending on the capacity and coverage of licensing restrictions, operators may use different levels of sharing. In addition, the agreement may not be restricted between two operators; an operator can have several agreements with other operators in different regions.

### 2.1. Passive Infrastructure Sharing

Passive infrastructure sharing is identified as the options available for mobile operators intending to share passive elements in their radio access network. Such sharing of radio sites, termed 'site sharing' or 'collocation' has become popular since the year 2000 [11]. Normally, the operators enter an agreement to share sites directly, but lately, there are enabling third parties involved in such agreements, which provides towers to the telecommunication operators. These so called 'tower companies' already have established footprint in mature markets where as they are coming up in the emerging markets as well, like India and the MENA region [20]. Generally, site sharing involves sharing of costs related to trading, leasing, acquisition of property items, contracts and technical facilities and the sharing of passive RAN infrastructure, i.e.,

- Masts and Pylons, electrical or fiber optic cables;
- Physical space on the ground, towers, roof tops and other premises,
- Power supply, air conditioning, alarm installations and other passive equipments;
- Protecting access;

Site sharing allows operators to reduce both capital (CAPEX) and operating (OPEX) expenditure by reducing their investments in passive network infrastructure and in network operating costs. Site acquisition costs and expenses for civil works account for up to 40% of the initial investment to the fixed assets. Within recurring costs, site-related costs typically make up 5-20% of network OPEX, with the bigger number applying for sites that are leased, not owned. The sharing of electrical equipment, such as air conditioning, further makes power consumptions an addressable cost item, which represents roughly 3% of the network OPEX [13]. It may also be a way of overcoming planning and other regulatory restrictions and to meet environmental concerns. Site sharing facilitates rollout, bringing more wireless services to low populated and rural areas. Because of the cost saving aspects, site sharing may also contribute to making wireless services more affordable.

Passive infrastructure sharing may be an effective option for upgrading second generation (2G) mobile services to third generation (3G) mobile communications and broadband wireless access technologies in emerging markets [20]. Operators that provide 2G mobile services may upgrade to 3G by collocating the required 3G equipment on their existing towers and masts. This may be a very cost effective option for operators, even if building a 3G mobile network would require a significantly larger number of sites. In countries with largely developed 2G networks, the collocation of 3G equipment on 2G infrastructure may provide a substantial advantage to incumbent 2G operators compared to new entrant 3G operators. This is mainly the case in Western European countries, where 2G networks have been deployed in a relatively early stage and where incumbent operators have gained a substantial advantage over newcomers. Therefore, non-discriminatory regulatory restrictions are imposed on existing 2G operators, requiring those to provide access to their facilities to new 3G operators under the same conditions as they provide to their own business. If new entrant 3G operators are not provided access under these conditions, they may not have a fair chance of competing on the 3G market with incumbent 2G operators. Other countries may be faced with this issue when 3G services are upgraded to 3.5 or 4G mobile services – especially where 2G infrastructure has been deployed later and 3G infrastructures is intended to be used as a predominant type of technology for mobile services.

There are different kinds of site sharing agreements that are considered by operators. Such agreements may be unilateral (one operator agrees to provide access to its facilities to another operator), bilateral (two operators agree to provide mutual access to facilities) or multilateral (involving several operators) [11]. In addition, such agreements may concern one individual site or be a framework agreement for several sites or for all the sites in a certain geographical region. Bilateral agreements for regional site sharing are particularly interesting for operators from an economical point of view. Operators agree to use each other's passive infrastructure in certain regions, to avoid having to build new masts or sites where they agree to share sites. This enables the operators to offer service coverage in a larger geographical area, which is particularly attractive for operators subject to geographic coverage obligations. Regional site sharing allows operators to save considerable OPEX and CAPEX and is an alternative to national roaming arrangements.

Most site sharing agreements do not restrict competition between operators; rather they generally allow operators to keep independent control of their respective networks and services. As a result, site sharing agreements generally do not lead to the harmonization of networks thereby making the service offerings and prices charged by the site sharing operators indistinguishable. Full competition is assured where operators retain independent control over their radio planning and the freedom to add sites, including non-shared sites. In that way, operators are free to increase their network capacity and coverage. Better coverage and capacity may be a competitive parameter, as operators may be able to distinguish themselves based on network quality and transmission capacity. It is also important that site sharing agreements do not contain exclusivity clauses, prohibiting operators from concluding similar agreements with third parties. Site sharing arrangements that fulfill these conditions are not likely to restrict competition between operators. Finally, site sharing agreements may have a positive impact on competition, since the savings achieved may eventually be passed on to consumers, increasing quality of service and decreasing price.

## 2.2. Active Infrastructure Sharing

The active sharing of facilities is an advanced technical model which involves a mutual sharing of not only passive, but also 'active elements' of the network that can be managed by the operators - installed in base stations and mobile network equipment, access node switches and finally the management systems of fiber optic networks. Additional savings on CAPEX and OPEX can be realized by sharing the active RAN infrastructure, i.e., BTS and BSC in 2G networks or NodeB in 3G networks. Active infrastructure sharing is more complex, since it covers the essential elements of value creation in the chain of economic activity. Many countries regulate the sharing of active infrastructure, fearing that the practice will promote anticompetitive behavior, such as agreements on price or service offerings. However, many regulatory authorities are taking more lenient approach to active sharing, as operators increasingly compete based on the quality of their services and not on the features of their networks. In this section, we look at the different options available for operators to perform active sharing of mobile networks.

### 2.2.1. Antenna sharing

This level is defined by the pooling, in addition to the passive elements of the site radio, antenna and all the associated connector (coupler, cable). In principle, antenna sharing can be considered as an extension of site sharing. In antenna sharing arrangements, operators may also share the TRX (transmitter and receiver), whereby demanding the sharing of the spectrum too. Although, spectrum sharing is technically possible, there are several licensing and regulatory challenges in particular because of spectrum regulations.

Antenna sharing has the potential to increase CAPEX and OPEX savings by operators as compared to simple site sharing. However, the amount of additional savings may be limited, since the additional costs of antennas and transmission equipment are relatively small. While it is technically possible for the operators using different sets of frequencies to share and antennas, it may not be an advisable option when the radio optimization strategies are not aligned between operators. The radio optimization strategy of the sharing operator governs the positioning of the antennas, which may introduce conflicts with the incumbent operator strategy. Currently, equipment manufacturers are supplying antennas which are adequate for antenna sharing [21].

### 2.2.2. Base Station sharing (Node B)

The base station (NodeB) is the device placed next to an antenna in the context of 3G networks. The base station contains a number of devices that are necessary to control the transmission and reception of signals. The sharing of base station is possible on condition that each operator:

- Retains control of the Node B "logic" so that it can operate independently of the partner operator frequencies that were assigned;
- Retains control of equipment assets of the base station such as TRX devices that are in charge of the transmission / reception on the radio channel

### 2.2.3. Sharing of Base Station Controller (RNC)

Sharing RNC is possible when it comes to maintaining control logic inside the RNC of each operator independently of one another. Figure 3 illustrates one RNC (Radio Network Controller) is physically divided into two or more logical RNC, belonging respectively to different operators. Each RNC has its own PLMN code logic, and its carrier frequencies. The RNC is physically separate but logically shared. This retention of control logic for each operator on the traffic allows the operator to ensure the proper control of this equipment. The operator thus control critical functions and operations performed by the RNC, including:

- Allocation and optimization of radio resource (admission control, allocation of spreading codes, power control, load control cells, the quality management of service)
- Mobility Management and Control hand-over settings.

RAN sharing solution enables operators to share their RAN acting logically as two separate access networks. This division uses two carriers and two MNCs operating in the same physical equipment. Some examples of implementation of RAN sharing appeared very recently. However, these examples are so far deployed only in the 2.1 GHz band to be used for serving low density areas with 3G [23]. Sharing model is not interesting in the densest areas since the income generated in these regions allows operators to deploy their individual networks.

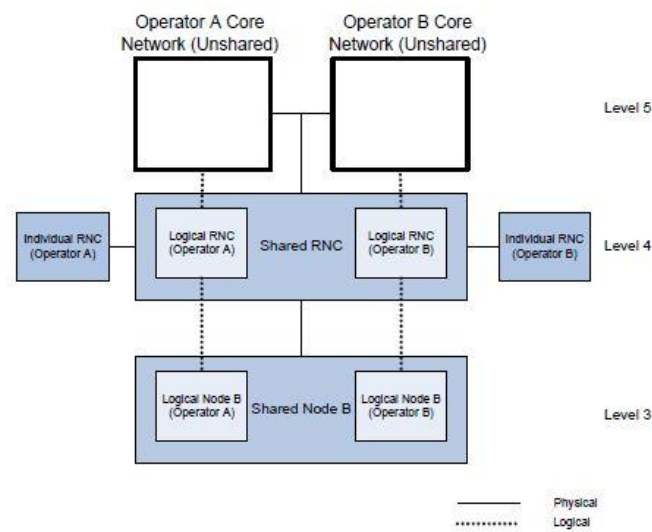

Figure 3: Schematic diagram for RNC sharing

In the case of full RAN sharing, regulations mandate generally that the shared elements are functionally separated, allowing the operators to retain independent control of all the parameters that are determinant of the quality of the network. This implies that the communication between the RNC and NodeB has to be under the independent control of one operator as far as its service is concerned. The operators must also independently control all the parameters that determine the quality of the network, such as coverage, speed and the handover parameters. In addition, operations, maintenance and network control must also be separated. Those elements may be under individual control by each operator or under joint control of an independent third party that operates the shared network on behalf of the sharing parties, also introducing Chinese walls between the sharing operators to enforce functional separation [14]. This independent third party may be a joint venture between the sharing parties or an outsourced service provider.

MORAN (Multi-Operator RAN) [17] is a technical solution where multiple virtual radio access network instances are implemented by splitting the BTS, BSC, Node B, and RNC into logically independent units being realized by a single physical instance. These virtual radio access networks are then connected to the respective operator core network – mobile switching center (MSC) and serving GPRS support node (SGSN) for circuit and packet switched traffic, respectively. Operators continue to use the dedicated frequency ranges that they were awarded by the licensing bodies, and broadcast their own individual network identifiers such that they maintain full independence in their roaming agreements and the sharing is not visible to their subscribers. With MORAN, all previously mentioned cost items are again addressable, but larger savings are obtained in various categories, like electrical power, and maintenance, because again the number of elements is reduced.

MOCN (Multi-Operator Core Network) is another active RAN sharing solution which has been defined in 3GPP Rel. 6 for 3G networks [19], where Node B and RNC are shared among multiple operators and frequencies are pooled. Addressable cost items are identical to MORAN, but while frequency pooling results in further marginal savings of equipment investment and equipment related costs due to a lower number of carrier units in extremely low-traffic areas, operators have to give up their independent control on traffic quality and capacity to a large extent. Subscribers using pre 3GPP Rel. 6 mobile terminals may realize that the network is shared. Under regulatory aspects, 3GPP MOCN's feature of frequency pooling may exclude the MOCN solution from being used in certain markets.

### 2.2.4. Core Network sharing

Core network sharing relates to the sharing of servers and the core network functionalities in addition to radio equipment. However, the core network performs several functionalities that are essential for the performance of an operator's service and contains a large amount of confidential information concerning the operator's business. Accordingly, it may be complicated for competing operators to share a core network. However, there are other varieties of sharing according to which operators may use the same core network to provide their services, such as national roaming, or through an MVNO construction. In addition, with the emergence of the so-called next-generation core networks in which the switching and the control and service functionality is physically separated, network sharing may move into the domain of core network switching while enabling service differentiation and confidentiality. The networks' "home" remain separate in core network sharing, which allows the differentiation of services. This is to pool the switches (MSC) and routers (SGSN) of the fixed network operator. A general schematic representation of core network sharing is shown in Figure 4.

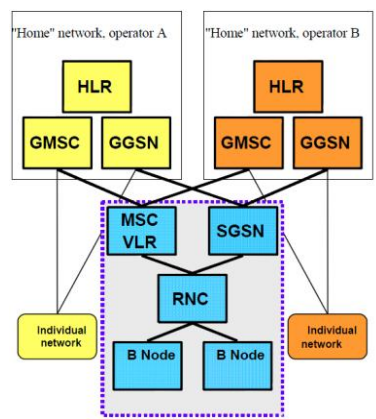

Figure 4: Schematic diagram for core network sharing

### 2.3. Roaming

National / International roaming is a form of sharing allowing customers of a mobile network operator to use mobile services when they are in an area not covered by its operator. From the beginning of 2G networks, roaming has always been employed as a means of virtually extending the geographic coverage of an operator by allowing its subscribers to use another operator's network. International roaming is the natural solution to serve one's customers abroad, where the operator has no license and no business. Roaming is also used on a domestic basis, as national roaming, typically to grant to a new entrant – or "greenfield" –operator nationwide coverage right from the start, when the operator rolls out its network initially in the urban and suburban areas and is not yet present in the rural areas. Roaming-based options in the context of network sharing, instead, mean that one operator relies on another operator's coverage for a certain, defined footprint on a permanent basis. Such dependence can be either unilateral or bilateral, regionally split or for the network as a whole.

A geographically separated network does not contain shared nodes. Each operator has its own carrier and its own MNC and builds its own access networks and core network and covers different areas of the country. There are three options for providing national coverage [16]. The first option is to sign a national roaming agreement, and to share the load. If the operators decide to retain dedicated independent core networks or only share the radio access network in a certain region, the "shared RAN with gateway core" solution can be deployed. Similar from a point of view of addressable cost items, compared to the active RAN sharing solutions outlined above, this approach, however, does not require specific features in the RAN equipment, as the sharing is fully implemented by roaming features that need to be implemented in the core network. The shared RAN is connected to the core networks of the sharing partners via a so-called gateway core consisting of MSC, SGSN, and visitor location register (VLR). In this solution, either frequencies are pooled, or only the frequency spectrum of one of the participating operators is used, such that there is no independent control of the traffic quality and capacity for the operators. If only one spectrum is used, capacity is substantially reduced; the pooling of frequencies is again subject to restrictive regulatory policies. Another solution is consistent with current releases in 3GPP, where both operators have their own access network, but a shared core network. In this case, the operators only retain that portion of the core network separate which also an MVNO (mobile virtual network operator) would own, i.e. home location register (HLR), authentication and billing system. In this case, all OPEX related costs are shared. The Figure 5 shows the three cases.

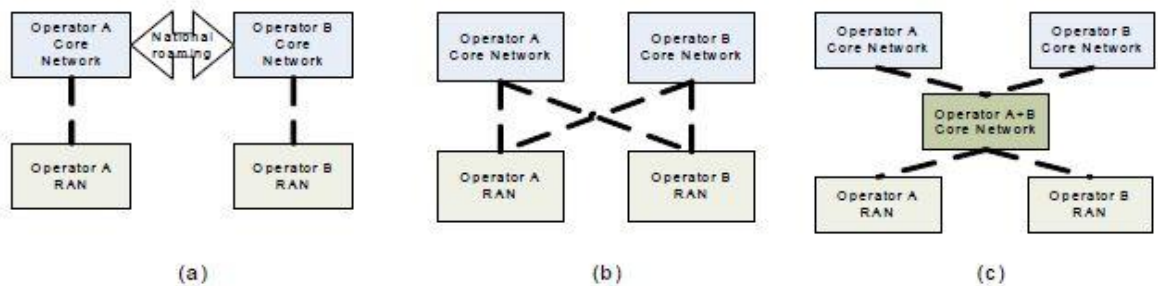

Figure 5: Three ways to connect networks that are geographically divided (A) Uses national roaming, (b) connects the two RAN with separate core networks (currently not standardized) and (c) uses a shared core network [2]

### 2.4. Mobile Virtual Network Operators (MVNOs)

A mobile virtual network operator (MVNO) is a virtual operator that provides public mobile telephone services without owning mobile frequencies or mobile access network, be it 2G or 3G networks. The MVNOs operates by reselling wholesale minutes that they have purchased from an existing infrastructure owner (a mobile network operator, or MNO). Most MVNOs have their own core network (including a billing and identification system) and only require access to the mobile operator's radio access network. From a technical standpoint, the MVNO subscriber is roaming on the permanent network of MNO. These operators are new players in the value chain of the mobile market. However, the degree of success of MVNOs and the business models adopted by them vary considerably from case to case and from country to country. In the UK, there are a number of successful MVNO's, the largest one being Virgin Mobile, with over 5 million subscribers in 2009. In the Netherlands, there are approximately 50 MVNOs and other service providers without a network, active in the market [22].

Although the presence of MVNOs may boost competition in certain markets, it is not a solution in markets where mobile networks have not been widely rolled out. MVNOs depend on the existence of previously deployed networks in order to provide their services. However, when operators do not utilize their full capacity, providing access to MVNOs may be a good alternative to bring more affordable services to the market. Entering into MVNO agreements may be commercially interesting when operators have spare capacity on their networks. Operators have an incentive to make this spare capacity available for alternative operators and boost their revenues. New entrants, possibly with a strong brand may want to enter the mobile market. Accordingly, regulators may consider facilitating the entry of MVNOs, whether or not regulated access, in order to boost competition and affordability of services.

### 2.5. Technical constraints of Infrastructure sharing

The sharing of network infrastructure requires coordination and cooperation between the involved network operators, with the increase in the level of sharing. Such cooperation shall bring forth multiple constraints on the activities of the concerned operators, which ultimately limits their flexibility of operation. These constraints particularly affect the operational elements in the deployment and operation of networks and can have an impact on the ability of operators to differentiate themselves in terms of services or quality of services. The following items are presented for analysing the technical constraints that would apply to the operators based on the expected level of sharing.

**Technical constraints related to passive sharing**

In the case of site sharing, the operators must take into account the following constraints:

- The qualifying sites to share (electromagnetic compatibility, models blankets, site area, optimizing 2G-3G)
- Installation of equipment on the shared site (access and safety, engineering site deployment schedule of operators)
- The operation and maintenance of equipment (on-site, monitoring and steering of networking equipment).

**Technical Constraints and drawbacks of active sharing**

- **Sharing of antennas:**

In case of antennas, one must consider additional constraints related to:

- The need for common choices, affecting the quality of service (technical diversity reception and transmission, radio planning, architecture of the antenna, use of TMA (Tower Mast Head Amplifier),
- The influence on the planning of the radio antenna amplifier linearity over several frequency bands,
- Taking into account in planning radio 3 dB loss induced by the coupling of the common antenna, for the separation of equipment connected to it.

- **Sharing NodeB**

In case of base stations (NodeB), we must take into account additional constraints related to:

- The use of NodeB containing at least two carriers (a significant difference between frequency bands of operators provides additional technical complexity),
- Limited number of operators (typically 3 or 4),
- A risk of lead single manufacturer solutions (in particular because of the interoperability links NodeB - RNC),
- Potential conflicts on the quality levels depending on the services available (power sharing),
- The operation and maintenance of shared assets.

- **Sharing the RNC**

In case of sharing of base stations controllers (RNC), we need to take into account the same types of constraints for sharing the Node B, which are still relevant in the case of the RNC, and additional constraints related to:

- Management of separation of the RNC functions (radio access configuration, performance management and quality of radio services),
- Interoperability between equipment from different manufacturers (hardware and software configuration),
- Interoperability between RNC and shared RNC, connected through the IuR interface [6] to ensure the handover (soft handover).

- **Core network sharing**

In case of equality of elements in the core network, it must take into account additional constraints related to:

- A choice of design of equipment common (NodeB, RNC, MSC, SGSN) to handle the traffic associated with the provision of services of each operator,
- A design package from core network management and service quality,
- The need to support intelligent network protocols consistent to ensure continuity of customer service of each operator when roaming on the shared network.

## 2.6. Outsourcing

In a context of growing competition in the market for telecommunications services, operators are seeking to become more flexible and increase their potential for future growth. For operators, outsourcing is a strategic and economically convenient choice. This enables them to focus on their core business, and to achieve by third party outsourcing service provider (including other operators, professional firms, manufacturers etc) all peripheral activities such as customer management and operation of telecommunication networks. These are secondary activities, according to the standards and specifications that they have chosen, and whose conduct of the operators is the responsibility of these stakeholders, for a fee.

To achieve this, the telcos have three models of collaboration as possible choice:

- The role of traditional traders that it focuses on the network infrastructure and delegate the marketing and sales, programming and packaging and advertising management functionalities to one or more third parties.
- The integrated operator which is becoming increasingly a provider of media and ensures full integration of the network operations business;
- Finally, the "new generation" operators who outsource network management to focus on the functions of billing and managing customers, marketing and sales, programming and packaging, and management advertising.

In the case of network sharing, instead of operating and maintaining the shared networks, the operators can out-task all network operations to 3rd party outsourcing service providers which is crucial for achieving higher operations related savings for all the operators involved in network sharing. The operations-related items in a shared network can simply be performed in a more cost-effective way by one single party, which achieves economies of scale, higher utilization of the field force, and less coordination costs. In the case of passive sharing, where only sites are shared, all the involved operators could in principle employ separate teams, even if awarding outsourcing contracts to a common provider of managed services (field services outsourcing including field services itself, spare logistics and repair, site maintenance, etc) would lead to significant cost cuts. But, in the case of maintenance of shared active equipments, it can be more efficiently carried out by a single party than having separate work force managing them. To avoid the potential principle-agent problem, setting up a joint venture is an ideal choice for the operators and outsourcing is the most attractive alternative (total operations outsourcing in this case, including management of network operations center, network planning and optimization etc.).

The main players in outsourcing are global infrastructure providers. Their services cover almost the entire business value chain of network operators. These actors are taking advantage of opportunities in issuing contracts to supply equipment in the long term and the creation of sustainable and predictable income from multi-year service contracts. The top five players in outsourcing are Alcatel-Lucent, Ericsson, Huawei, Motorola and Nokia Siemens Networks (NSN) [8].

## 3. Economic Dimensions of Network Sharing

Infrastructure sharing between operators undoubtedly leads to a reduction of the investment made by each operator involved in the network sharing process. Network sharing may further help operators in attaining better coverage, since they may choose to use only those sites that provide deeper and better coverage, decommissioning sites with poor coverage possibilities. Operators may reinvest those savings in upgrading their networks and providing better roll out and coverage to end-users. Network sharing agreements may also bring substantial environmental benefits, by reducing the number of sites and improving the landscape. Schematically, the investments for the deployment of a 3G network can be divided as shown in Figure 6.

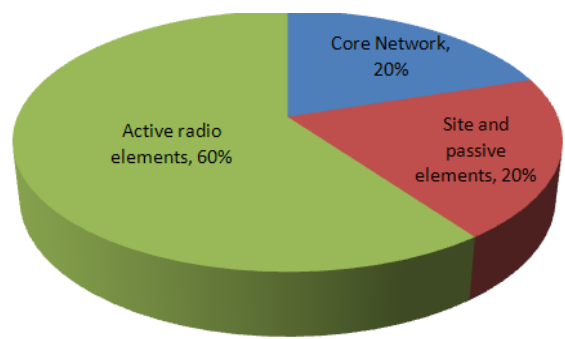

Figure 6: The investment allocation of network deployment

The sharing of sites and antennas, a combination of level one and level two sharing, can reduce on an average 20-30% of CAPEX costs. If the operators also share the radio network, there can be more savings, whereby the operators can save between 25 and 45%. Finally, the sharing of all the assets would decrease CAPEX by an additional 10%.

Promoting network sharing is a useful tool to encourage network deployment and coverage improvement in developing and emerging markets and un-served or under-served areas. There are several instruments that can be used to promote network sharing. National roaming arrangements are probably the most simple and effective arrangements. While national roaming leads to a certain level of uniformity between operators, it is important to analyze to what extent this uniformity leads to a significant restriction of competition. National authorities that have anti-competitive concerns may allow network sharing for a limited period (for example for a period of one or two years) of time in order to promote roll out in an initial phase of network deployment. After such an initial phase, operators would be required to provide coverage using their own networks.

Other types of arrangements, such as active infrastructure sharing, open access model (allowing and promoting the entry of MVNO's) and functional separation, may also work well to promote roll-out of wireless infrastructure and the advancement of competition. However, these types of arrangements may be difficult to enforce. Such measures require a strong regulatory system in the emerging markets, with appropriate powers to impose necessary measures. For example, it is relevant to note that a number of network sharing agreements in developed countries have preceded later mergers between the companies involved. In other cases, the companies involved in network sharing arrangements have not merged between themselves, but had to witness a consolidation wave taking place in the market. It is possible that this consolidation wave could have been avoided if operators would have been allowed more freedom to share their infrastructure and to concentrate on competing as far as their services are concerned.

The magnitude of the economical aspects of infrastructure sharing is difficult to assess, since it heavily depend on the particular level of sharing and geographic deployment strategy chosen [11]. Operators should carefully consider the sharing model they intend to adopt. Depending on the model chosen, the opportunities for competition differ. There are also a number of obstacles to be overcome when dealing with network sharing agreements. From an economical and practical point of view, network sharing is a large and complex process that requires a number of managerial resources. Sharing constraints vary from a country to another with major components differing even within a given country. Cost of passive items can largely vary according the country (power, site rental, taxes) and site constructions can incur variable costs. Moreover, the transport networks are of different types and have variable costs associated to them. All these factors account for the difficulty in identifying a generic economic case for infrastructure sharing. Nevertheless, we identify the emerging tendencies regarding CAPEX or OPEX repartition from a macro point of view in developing markets.

### 3.1. CAPEX/OPEX analysis

Cost saving is the main driver when considering infrastructure sharing. The following analysis give an idea of what the most important items for CAPEX cost savings are for a Mobile Operator, both for developed markets and for emerging ones [22]. The data below is an average observed values.

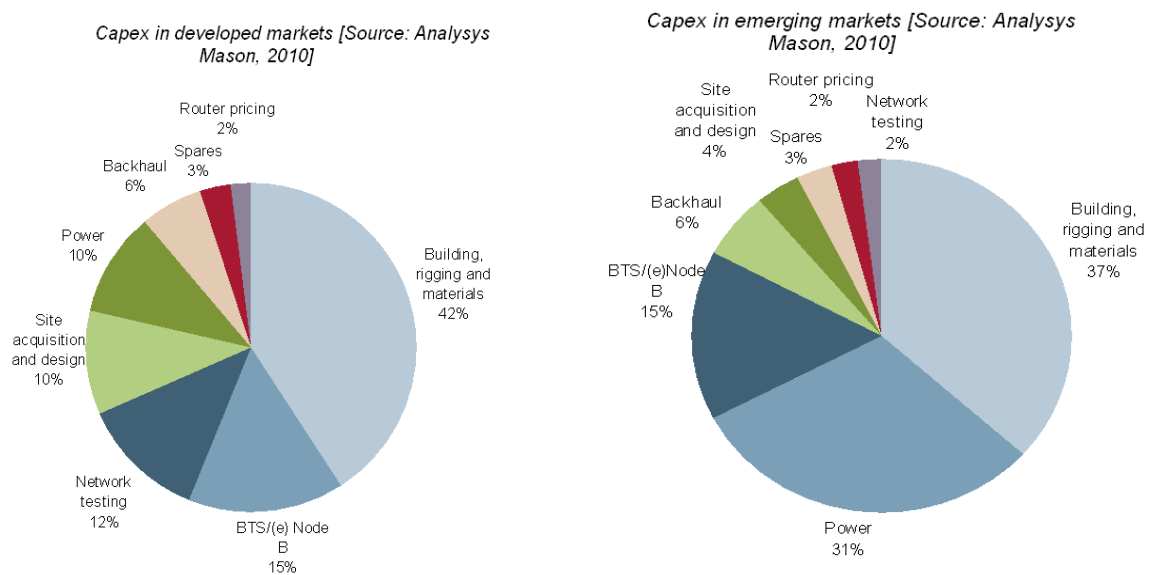

Figure 7: CAPEX analysis for infrastructure costs in developed and emerging markets

From Figure 7, we can observe that the pertinent CAPEX items for sharing are not similar for developed and emerging markets. This largely influences the sharing models for emerging markets when compared to the models in developed markets. In emerging countries, the 3 most important items which can be shared represent 87 % of the costs are civil and site acquisition and design (41%), power (31%), and BTS/NodeB (15%). Compared to that, in developed markets, the most important cost item is civil and site acquisition costs which amounts to 52% of the cost, where as other costs turn to less important. The power becomes clearly the main cost item in emerging countries. The access to electrical network is difficult and its coverage is weaker compared to developed countries. As a conclusion, infrastructure sharing is the most interesting choice in terms of improving CAPEX costs for new entrant operators in emerging markets. Passive sharing and maintenance will be ideal choice with joint ventures for outsourcing the site maintenance to local collaborators which can considerably reduce the CAPEX costs for new entrant MNOs in emerging markets.

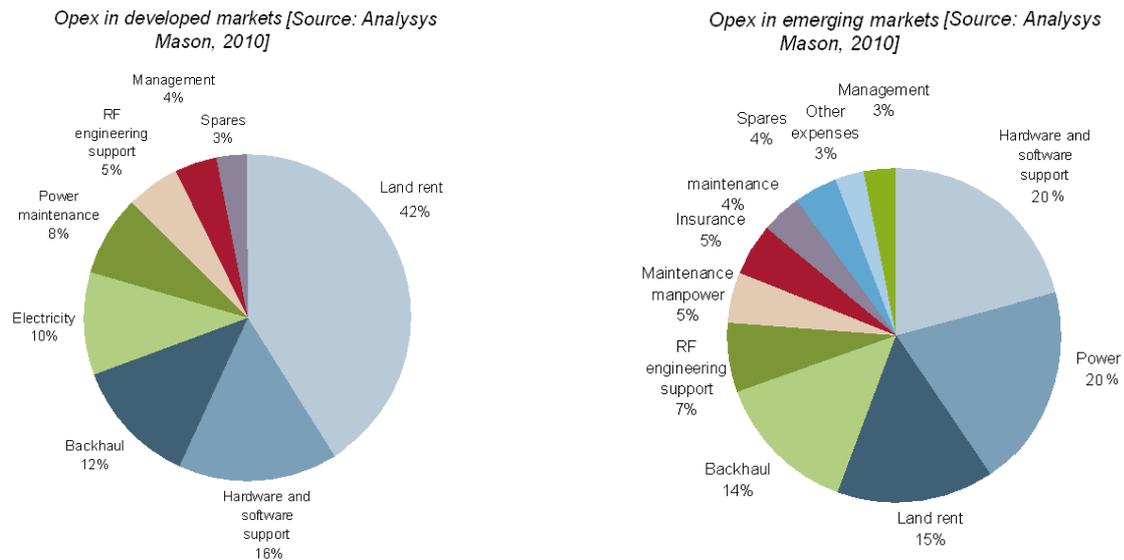

Figure 8: OPEX analysis for network operation costs in emerging and developed markets

Similar to CAPEX analysis, the following Figure 8 [22] gives an idea of what the most important items for OPEX cost savings are for a mobile operator, both for developed and emerging markets.

In emerging countries, the 4 most important items which can be the sharing represent 69 % and are hardware and software support (20%), power (20%), land rent (15%) and backhaul (14%). The OPEX is more shared between different OPEX in developing countries, whereas in developed countries the land rent (site) represents 42 % of the OPEX. That explains why the developed countries try to share the site so as to reduce OPEX. From the analysis of CAPEX and OPEX costs, the sharing items where costs saving (CAPEX and OPEX) can be achieved in a network in emerging countries are at the site (Civil and Site Engineering, Renting), power (Electricity, Diesel and Solar), RAN with BTS/eNodeB (Hardware and Software), and backhaul.

Where passive sharing (mainly through site sharing) is largely used particularly in emerging countries, Active sharing is however even more complex to achieve owing to the choice and complexity of the technical solutions (to find an agreement about infrastructure manufacturers technologies, frequencies) and the complexity associated to operating a shared RAN network (to find an agreement about network design, radio optimization, software level, release level, QoS etc.). Careful negotiations with relative simplicity, trust and transparency are important goals to achieve agreements for RAN sharing. As an example, RAN sharing is realized in Spain between the operators, Orange and Vodafone [16].

### 3.2. Emerging countries characteristics

Some general recommendations for infrastructure sharing in emerging countries are drawn as discussed in the following:

- Rural sharing is strongly recommended (both 2G and 3G),
- Sub-urban sharing could be recommended in some cases,
- Urban sharing is not recommended for 2G, where as 3G sharing could be recommended in some cases,
- Co-locate 3G sites with existing 2G infrastructure sites.

However, a lot of parameters shall be taken into account. Most of these parameters are depending on the local situation and constraints. Therefore, there is no generic case: practically, each case needs to be considered as a specific situation.

In the case of existing networks in emerging countries, some constraints have to be analyzed as described in the Table 1.

| Sharing domain | Constraints |
|---|---|
| Site | - Area of the site (is the site sufficient for new equipments or additional site area need to be acquired) |
|        | ▪ Mast of site (dimensioned enough to receive new antennas)                                                                                                                                                                                                                                                          |
| Energy | ▪ Generally, energy is dimensioned just for the current needs (necessary to adapt the energy to the new requirements)<br>– Electrical: change the standing charge<br>– Battery and Emergency Energy: to add new batteries<br>– Diesel: change the existing generators by new generators<br>– Solar: to add new solar panels (site extensions requirements) |
| RAN    | ▪ Add new radio components to meet additional traffic demands<br>▪ As the network is already operating, it is necessary to accept the constraints of the other operator: network design, radio optimization, software level, quality of service.                                                                     |
| Backhaul | ▪ Microwave (dimensioned enough to receive of adding traffic)<br>▪ Capacity, Lines                                                                                                                                                                                                                                 |

Table 1: Constraints analysis in the case of existing networks in emerging markets

In the case of new networks, some constraints have to be analyzed as discussed in the Table 2.

| Sharing domain | Constraints |
| --- | --- |
| Site | ▪ Choice of site or geographical splitting<br>▪ Number of sites (Coverage quality) |
| Energy | ▪ Choice of power: electrical if possible, diesel, solar |
| RAN | ▪ Technical complexity: to find an agreement about infrastructure manufacturers, technologies, frequencies.<br>▪ Operationally complex: to find an agreement about network design, radio optimization, software level, release level, QoS etc |
| Backhaul | ▪ Choice of the type of backhaul: lines, Microwave, VSAT (Very Small Aperture Terminals)<br>▪ Data traffic |

Table 2: Constraints analysis in the case of new network installations in emerging markets

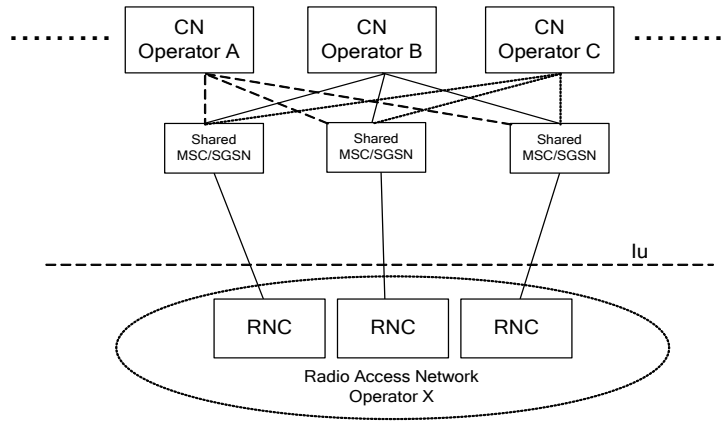

Figure 9: Configuration for GWCN sharing. In addition to RAN sharing, operators also share core network [6]

The sharing of infrastructure between operators has the advantage of promoting better timing of investments, thus helping to reduce the initial financial burden of operators. However, it does not per se resolve all problems related to the financial viability of projects and operators. Some regulators restricted infrastructure sharing to level 4 as defined above, and they ruled out the level 5 using the criterion that each operator must maintain control of its network "logic" of its own data traffic and the use of its own frequencies, to control all the components that determine the quality of their network and service provision.

From a consumer perspective, if this "concentration" in terms of infrastructure may result in maintaining or increasing self-service offerings (MVNO agreements), the "focus" on infrastructure is not incompatible with maintaining competitive benefits to consumers. A regulatory framework is necessary, but it must strike the right balance. And to support the phenomenal rate of growth of their clients and maintain a remarkable margin despite a lower ARPU, operators began to opt for outsourcing the design, deployment and management of networks and also manage subscriber data and Services. This trend practiced in recent years in favour of suppliers limits the role of the operator to acquire and manage customer relationships. In order to benefit mutually from the potential benefits of network sharing, the establishment of a rigorous integration and an operating plan implemented by a neutral party, provides a solid foundation for collaboration and allows a clear governance principles that dictates the decision-making processes for the efficient processing of CAPEX and OPEX of allowing each operator to choose its own strategy internally within the collaboration.

## 4. Practical use case

The objective of this section is to evaluate the CAPEX/OPEX saving obtained thanks to the different infrastructure sharing models (MOCN, MOCN + backhaul sharing, MOCN without spectrum sharing (MOCN – spectrum), GWCN, GWCN + backhaul sharing, GWCN without spectrum sharing). The Table 3 below illustrates the different sharing levels and the associated configurations.

| Shared Elements | MOCN | MOCN + Backhaul | MOCN - Spectrum | GWCN | GWCN + Backhaul | GWCN - Spectrum |
|---|---|---|---|---|---|---|
| Passive infrastructure | X | X | X | X | X | X |
| NodeB | X | X | X | X | X | X |
| RNC | X | X | X | X | X | X |
| Backhaul | - | X | X | - | X | X |
| Spectrum | X | X | - | X | X | - |
| SGSN | - | - | - | X | X | X |

Table 3. Infrastructure sharing configuration

### 4.1. General hypothesis

The main guidelines of the proposed study are as in the following:

- The country is divided into 3 independent regions; urban, sub-urban and rural with the same number of users and service profiles. The traffic from each region will be associated to an independent RNC.
- We use the existing GSM sites to build 3G network (the cost of site building is out of the scope of this study).
- The amortization period is assumed to be 5 years.

Table 4 provides the specific parameters for the areas/sites considered by this study[1]:

| Areas | Urban | Suburban | Rural |
|---|---|---|---|
| NodeB | 78 | 58 | 108 |
| Subscribers | 17 700 | 17 700 | 17 700 |
| CORE Network | 1RNC 1SGSN/1GGSN | 1RNC 1SGSN/1GGSN | 1RNC 1SGSN/1GGSN |

Table 4. Network and services configuration description hypothesis

The per-segment OPEX/CAPEX costs repartition is illustrated in Figure 10 and Figure 11 (without sharing). Figure 10 depicts the per area CAPEX repartition. The Core Network (SGSN+GGSN) as well as the O&M present the smallest part of costs (between 8% and 17%). The Backhaul is the most important one, ranging from 32% for suburban area to 41% for rural area. The RNC cost is close to the Backhaul cost excepted for rural area where it dropped to 11%. The NodeB cost varies from 23% for Suburban area to 29% for Rural area.

---

[1] The coverage, topology and the density is not the same in urban, suburban and rural areas. Thus, the deployment parameters and service configurations are different. The numbers in Table 4 are considered so as to provide a fair comparison between the scenarios in the case of the same number of subscribers.

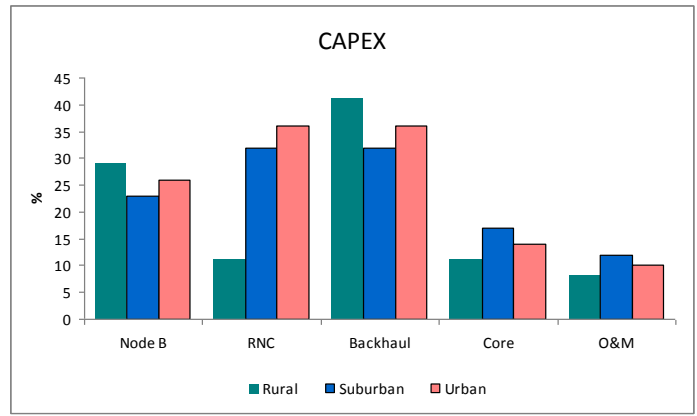

Figure 10: CAPEX repartition

From OPEX perspectives, it is unsurprisingly, the international connectivity (e.g., submarine cable cost, which are a periodic expense accounted as OPEX costs) which presents the most important part of the cost (more the 50%), then the UTRAN (RNC + NodeB). The Licence and Core Network present together around 10% of the total OPEX costs.

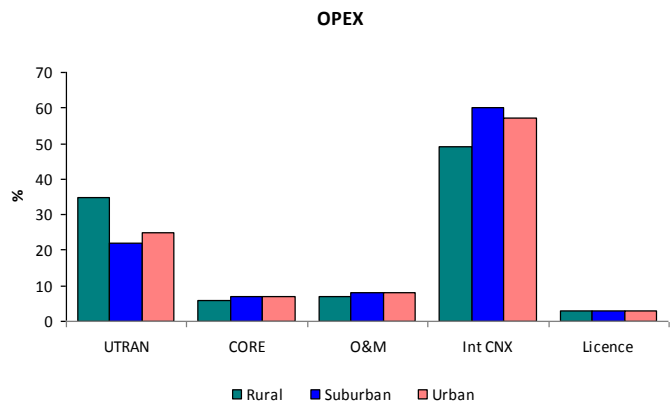

Figure 11: OPEX repartition

In order to estimate the gain related to the introduction of infrastructure sharing techniques, we consider the inputs: Network sharing solution (MOCN, GWCN, etc.) and the type of areas. On the output, we obtain the cumulative CAPEX/OPEX and the respective savings. We assume that sharing any equipment/segment will bring 50% of CAPEX/OPEX saving.

#### 4.1.1. Results

The source of CAPEX savings are the lesser radio and transmission equipments and the existing network whereas the source of OPEX saving are the avoidance of site rent, lower site equipment maintenance and less network Staff required. Overall, based on our study, it is the GWCN + backhaul sharing which provides the best gain in performance for all areas, from both OPEX and CAPEX perspectives. This is due to (valid for all configurations) the fact that such configuration allow maximum network element sharing (NodeB + RNC + SGSN + Backhaul) and can operate in both dedicated and shared spectrum. Due to its high cost (OPEX part), we also notice that significant saving can be brought if the international connectivity is shared (we consider the following hypothesis: 3G clients' ratio of each operator as 50%; the sharing/splitting ratio for each operator is of 50%).

#### 4.1.2. Urban dense Area

For dense area, with a total of around 27.12% saving (both CAPEX/OPEX included), the GWCN+ Backhaul sharing is the best sharing approach for dense areas. Sharing the spectrum brings 1% additional savings (GWCN + Backhaul Sharing and Spectrum Sharing). GWCN + Backhaul Sharing surpass MOCN + Backhaul Sharing by 1.72%, in the case of dedicated frequency. Depending on the approach, the achievable CAPEX savings vary from 25% to 48% whereas the achievable OPEX savings vary from 16% ~ 18%.

As mentioned earlier, GWCN + Backhaul Sharing offers the best OPEX and CAPEX reduction performance. Whereas the CAPEX saving vary considerably from an approach to another, the main difference in the OPEX saving reside in the spectrum sharing. In fact, sharing the spectrum provide 2% additional OPEX saving.

### 4.1.3. Sub-Urban Area

We observe the same results for suburban area, with a total of 25% saving (both CAPEX/OPEX included), the GWCN+ Backhaul sharing is the best sharing approach for dense areas. Sharing the spectrum brings 1.07% additional saving (GWCN + Backhaul Sharing and Spectrum Sharing). GWCN + Backhaul Sharing surpass the MOCN + Backhaul Sharing by 1.9%, in the case of dedicated frequency. Depending on the approach, the achievable CAPEX savings vary from 25.5% to 47.9%. Whereas, the achievable OPEX saving vary from 14.9% to 16.5%. Unsurprisingly, GWCN + Backhaul Sharing offer the best OPEX and CAPEX reduction performance. Whereas the CAPEX saving vary considerably from an approach to another, the main difference in the OPEX saving reside in the spectrum sharing. In fact, sharing the Spectrum provide 1.56% additional OPEX saving

### 4.1.4. Rural Area

It's the GWCN + Backhaul sharing which achieves the best performance even for rural area. Sharing the spectrum brings 0.7% additional saving (GWCN + Backhaul Sharing and Spectrum Sharing). We notice that the GWCN + Backhaul sharing outperform the MOCN + Backhaul by only 1.5%, in the case of dedicated frequencies. Compared to the Sub-Urban and Urban Area, a slightly higher OPEX and CAPEX saving is achievable. Depending on the approach, the CAPEX savings vary from 29.4% to 48.6%. Whereas, the achievable OPEX saving vary from 18.5% to 19.9%.

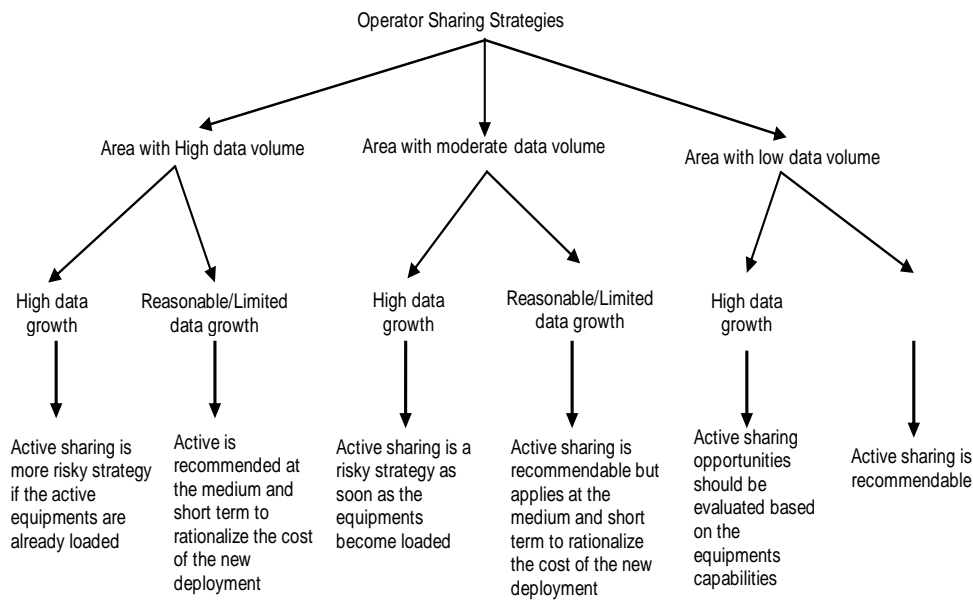

Figure 12: Infrastructure sharing recommendations

### 4.2. Lessons learned and Recommendation

The main lessons learned from the conducted study are the following (the Figure 12 summarises our analysis for the short term):

- GWCN + backhaul provide the best saving for all the considered cases (Dense, Sub-urban and Rural). The MOCN + Backhaul is the next option with slightly less achievable savings.

- International connectivity presents more than 50% of the global OPEX but is expected to decrease in the next few years, thanks to submarine cables and Pan-African fibre networks. An effective access and backhaul active sharing will be more beneficial.

- Sharing the active infrastructure implies sharing of the market throughout the sharing period. It is not effective as soon as the network becomes over-used or if the market is growing quickly. But it may make sense to look at advanced ways to share network in the long term. On one side, the operators will expect progressively new features for network differentiation (e.g., service priorities) and dynamic sharing policies in order to be able modify rules in case of changes in the forecasts. On the other side, operators would pursue to cooperate (share from time to time) the resources of the backhaul in order to achieve low-cost resiliency mechanisms: rerouting temporarily traffic to other operator links only in case of failure (or energy saving policy), due to the over dimensioning of backhaul links, sharing, and service priorities.

# 5. Evolution toward the LTE

For next generation of mobile network, LTE, 3GPP took into account network sharing in its specification. It has defined the MOCN [17] and GWCN (Gateway Core Network) approaches for the eUTRAN sharing. The Table 5 provides a high level comparison of the two approaches. In the MOCN approach, the shared eUTRAN is connected to several CN via the S1 interface. Each mobile network operator has its own EPC. Thus the MME, the SGW and the PGW are not shared and are located in the different CN. The S1 flex allows the eNodeB to be connected to the different CN. It also allows connecting the eNodeB to several MME and SGW in a given CN. Thus, allowing load balancing to be supported between MME and SGW of a given CN. In the GWCN approach (as shown in Figure 9), contrary to the MOCN approach, the MME is also shared between the different mobile network operators.

|  | MOCN | GWCN | Remarks |
|---|---|---|---|
| Internetworking with legacy networks | + | - | To support inter-RAT mobility, MME needs interfaces with legacy networks (i.e., SGSN). Sharing the MME leads to a tighter integration between the shared eUTRAN and each core network operator |
| Support of voice service with CS fallback | + | - | CS fallback need the support of the SGs interface between MMEs and the MSCs. Sharing the MME leads to a tighter integration between the shared eUTRAN and each core network operator |
| Support of voice services with IMS | = | = | Support of IMS is the best and future solution for voice over LTE |
| Support of roaming | + | - | In roaming MME in visited network needs to interact with HSS in home network. Having a shared MME is a drawback as HSS address of each roaming partner needs to be define in shared MME for each CN connected to the shared eUTRAN |
| Cost | - | + | Sharing the MME shares the cost. However, this depends on the context |

Table 5: high level comparison of MOCN and GWCN approaches for RAN sharing in LTE networks

# 6. Regulatory aspects

From an economical and practical point of view, network sharing is a large and complex process that requires a number of managerial resources. Therefore, the concrete benefits generated by network sharing should be analyzed by regulators and policy makers on a case-by case basis, taking into account the specific characteristics of each market involved. Promoting network sharing is a useful tool for regulators and policy makers to encourage network deployment and coverage improvement in un-served or underserved areas. There are several instruments that can be used to promote network sharing. National roaming arrangements are probably the most simple and effective arrangements. While national roaming leads to a certain level of uniformity between operators, it is important to analyze to what extent this uniformity leads to a significant restriction of competition. National authorities that have anti-competitive concerns may allow network sharing for a limited period (for example, for a period of one or two years) of time in order to promote roll out in an initial phase of network deployment. After such an initial phase, operators would be required to provide coverage using their own networks. Active infrastructure sharing, an open access model (allowing and promoting the entry of MVNO's) and functional separation may also work well to promote roll-out of wireless infrastructure and improving competition. However, these types of arrangements may be difficult to enforce. Such measures require a strong regulator and an efficient judicial system, with appropriate powers to impose the necessary measures. If operators are allowed more freedom to share their infrastructure and to concentrate on competing as far as their services are concerned, the large consolidation wave in the telecommunications market across the globe could be restricted.

# 7. Conclusions

Infrastructure sharing solutions have proven to be a critical lever contributing to the growth of the telecommunication sector. In this paper, we described the technological approaches that appear viable from today's perspective, considering current technology, and showed how to align these concepts with business and economic strategies, particularly focusing on emerging mobile communication markets. Both fixed and mobile operators should consider network sharing as a medium to cut operational and capital costs, and to focus more attention on innovation and differentiation in customer-facing activities. We also discussed the economic impacts of the various options on operational and capital expenditures of the operators. In the longer term, incumbent operators could leverage network sharing as a means for continued growth by structurally separating all or part of their network assets or spinning out network provider companies in competitive markets. We also evaluated case studies for the CAPEX/OPEX savings for different areas (including dense, suburban and rural scenarios) comparing different infrastructure sharing models (MOCN, MOCN + backhaul sharing, MOCN without spectrum sharing (MOCN – spectrum), GWCN, GWCN + backhaul sharing, GWCN without spectrum sharing). The main lessons learned from these studies were reported.

Finally, this paper has identified a number of best practices to promote competitive passive and active mobile infrastructure sharing:

- Establish clear, objective and transparent policy goals involving network sharing
- Establish clear guidelines for the conclusion of voluntary sharing agreements, including time limits to conclude agreements and to provide actual access.
- Create efficient dispute settlement mechanisms and judicial review, including specialized dispute settlement bodies.
- Allow and stimulate self regulation.
- Consider network sharing, in particular site sharing and national roaming, in rural and remote areas.
- Make thorough and objective assessment of the competitive situation, including research on consumer preference and consumer choice.
- Consider whether an open access model (such as the entry of MVNOs) or even functional separation would be viable, depending on the actual situation
- Regulators and policy makers should consider providing subsidies related to network sharing in rural and remote areas calculated to cover real costs and distributed in a competitive fashion.